\documentclass[twocolumn,showpacs,preprintnumbers,amsmath,amssymb]{revtex4}
\usepackage{graphicx}% Include figure files
\usepackage{dcolumn}% Align table columns on decimal point
\usepackage{bm}% bold math

\usepackage{psfrag}

\begin{document}

%\preprint{}

\title{A Theoretical Study on an Optical Switch Using Interfered
Evanescent Light}

\author{Naofumi KITSUNEZAKI}
 \email{kitsunezaki@it.aoyama.ac.jp}
\author{Jun-ichi MIZUSAWA}%
 \email{mizu@it.aoyama.ac.jp}
\author{Akio KITSUNEZAKI}
 \email{kitsune@it.aoyama.ac.jp}
\affiliation{%
College of Science and Engineering, Aoyama Gakuin University
}%

\date{\today}% It is always \today, today,

\begin{abstract}
 In an optical configuration consisting of a flat plate of vacuum
 between upper and lower spaces of uniform dielectric regions of $n>1$, 
 we have calculated two output light intensities for two input lights 
 from the Maxwell's equations 
 as functions of the incision angle, a light intensity ratio, 
 a phase difference of the two input lights, 
 and a thickness of the vacuum layer, 
 where the two input lights come from upper and lower dielectric regions
 with the same incision angles, 
 and 
 one of the output light goes into upper dielectric and the other goes
 into lower dielectric.
 We have found that, when evanescent lights exist at the upper and lower
 boundary and interfere each other, 
 there is one set of incision angles and phase differences for any
 combination of an input light ratio and a thickness of the vacuum layer
 where one of output lights becomes zero. 
 This finding will possibly lead to an innovative optical switch with
 which an optical output light can be switched on and off with a control
 light with an intensity much lower than that of the output light.
\end{abstract}

%42.79.Ta: optical computers, logic elements, inter-connectors, switches;
%          neural network
%42.25.Hz: Interference
%51.70.+f: Optical and dielectric property
\maketitle

\section{Introduction}
In the present wired tele-communication, 
the optical fiber using an infrared light with wavelength around 1500 nm
is playing the main role, 
but the signal processing at both ends of the transmission optical fiber
is done with electronics. 
That means there are optoelectronic and electro-optic converter
circuits between optical fibers and electronic circuits. 
In order to avoid a demerit of electronic circuits that they are
vulnerable of electro-magnetic noise and also to simplify devices by
eliminating optoelectronic and electro-optic converters, 
a purely optical signal processing device represented by optical 
switches has long been desired.

Optical waveguides are utilized in optical communication
devices as star-couplers and array waveguide gratings (AWG)
to add or to divide optical signals. 
A star-coupler is a device to divide the energy of an
optical signal carried by the core of an optical fiber into several
output optical signals. 
The AWG is used for a filter using a characteristic that multiple
optical signals interfere each other in a small space of a optical
waveguide circuit. 

\begin{figure}[tb]
 \begin{center}
  \includegraphics[width=60mm]{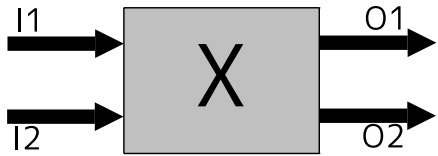}
  \caption{An optical switch of the present article}
  \label{fig:switch}
 \end{center}
\end{figure}

An optical splitter and an optical coupler are made of
an optical fiber or an optical waveguide. 
They can be the same device by exchanging their inputs and outputs
of lights.
To divide or add optical signals, 
the device can be an optical waveguide with a geometrical branch
shape or two parallel optical fibers. 
When two optical fibers are located in parallel and very close to
each other, 
the energy of an optical signal is transferred from an optical
fiber to other fiber. % by a kind of the tunneling effect. 
As the optical switching elements, 
semi-conductors or ceramics having characteristics of changing their
optical properties such as refractive index with electric field,
magnetic field, or temperature have been developed and utilized. 
Such devices as the one composed of a lattice of optical 
waveguides with optical switching materials buried at each
cross point of the lattice are proposed and some of them have been
realized. 

\begin{figure}[bt]
 \begin{center}
  \includegraphics[width=80mm]{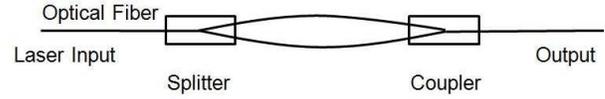}
  \caption{Mach-Zehnder Circuit}\label{fig:m1}
 \end{center}
\end{figure}

The evanescent light is a kind of light existing on places such as
the back side of a total reflection prism within a very small area, 
typically within a distance of one wavelength. 
The evanescent light can be developed from the Maxwell's equations, 
but it is only a decade ago that the evanescent light became one of
major research subjects. 
The evanescent light has already been utilized probably based on the
empirical findings. 
That is the divider or coupler made of two parallel optical fibers
mentioned above. 
The evanescent light appears on the surface of the input optical fiber
and the energy of light moves into the output optical fiber running
close by and in parallel. 
The amount of light energy transferred to the output fiber depends on
the gap between two fibers and the length of these two fibers running
together in parallel. 
The ratio of light energy splitting/coupling can be controlled in the
manufacturing. 
This type of a coupler/splitter is produced with carefully
adjusting the gap and length watching the intensity of output light. 
The produced coupler/splitter has two input optical fibers and two
output fibers, and that is one of the simplest example of the four
terminal circuits as shown in Fig.\ref{fig:switch}.

An example of a coupler/splitter application is the Mach-Zehnder
interferometer which is composed of a 50:50 splitter and a 50:50 coupler
with two transmission lines between them, as shown in Fig.\ref{fig:m1}. 
The splitter and coupler in this case are the same configuration as the
four terminal circuit explained above but one of four terminals is
neglected or the ratio of light energy to one of four terminals is
designed to be zero.~
The Mach-Zehnder interferometer can also be made by dielectric optical
waveguides.

A coherent light injected to the input of the Mach-Zehnder
interferometer is divided equally to two transmission lines and when
the two transmission lines are same, 
lights from the two transmission lines are simply added at the coupler
and the same light as the input light goes out. 
When there occurs a phase difference between two transmission lines
between the splitter and the coupler, 
sum of two lights having phase difference is given to the output
terminal. 
If the phase difference is $\pi$, then the output is zero. 
Therefore, a light switch or a light modulator can be made by using a
light phase controller and a Mach-Zehnder interferometer.

In this paper, we propose a new type of optical switch which is
expressed by a four terminal circuit as Fig.\ref{fig:switch}. 
Two lights having a particular relation between them injected into
two input terminals produce their evanescent lights and through
interference of the evanescent lights in this new device, 
the two output lights are switched on and off. 
We have theoretically studied the mechanism by solving the Maxwell's
equations including evanescent lights.

\begin{figure}[bt]
 \begin{center}
  \includegraphics[width=60mm]{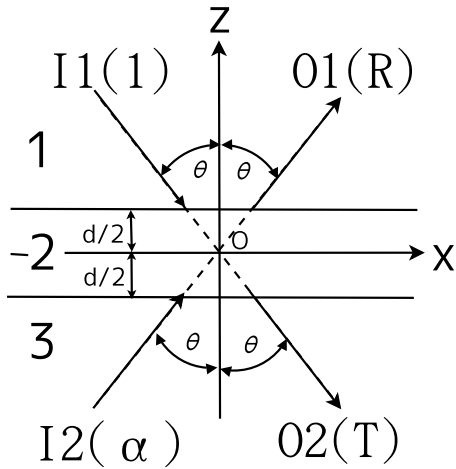}
  \caption{A theoretical model}
  \label{fig:Model}
 \end{center}
\end{figure}

\section{Calculation model and contents of this paper}
As shown in Fig.\ref{fig:Model}, 
we set a calculation model consisting of three regions. 
Region 1 : $z>d/2$, refractive index $n$,  
region 2 : $-d/2<z<d/2$, vacuum (refractive index 1), and 
region 3 : $z<-d/2$, refractive index $n$. 
A light (I1) with a vacuum wave length $\lambda$ and intensity 1 is
injected from region 1 with an injection angle $\theta$, 
and another light (I2) of same wave length with intensity $\alpha$ is
injected from region 3 with the same injection angle $\theta$. 
I2 has a phase difference (delay) $\eta$ relative to I1 when the two
lights arrive at the boundary at the same $x$.

In section \ref{sec:refraction}, we have calculated 
the ratio $\overline{R}$ of an output light (O1) intensity to 
an input light (I1) intensity, both in the region 1, with $\theta$
less than the critical angle where refraction lights propagate in
region 2. 
We have also calculated $\overline{R}$ with $\theta$ larger than the
critical angle in section \ref{sec:evanescent} taking the evanescent
light in region 2 into account. 
In section \ref{sec:parameter_condition}, 
we have discussed the conditions where $\overline{R}=0$ based on the
results attained in sections \ref{sec:refraction} and \ref{sec:evanescent}. 
The conditions for $\overline{T}=0$, the condition for the output light
into the region 3 being zero, is not discussed because the total
energy of the input lights (I1 + I2) is conserved to that of the
output lights (O1 + O2) in the present model, 
and the condition is clearly $\overline{R}+\overline{T}=1+\alpha$.

\section{Output intensity when the injection angle is less than the
 critical angle}\label{sec:refraction}
Because no electric charge and no electric current exist in the model of
Fig.\ref{fig:Model}, the Maxwell's equations to be solved are:
\begin{subequations}
\begin{align}
 \nabla\times{\bf E}+\partial_t{\bf B}&={\bf 0},\label{eqn:Faraday}\\
 \nabla\cdot {\bf B}&=0,\label{eqn:no-monopole}\\
 \nabla\times{\bf H}-\partial_t{\bf D}&={\bf 0},\\
 \nabla\cdot{\bf D}&=0,
\end{align}
 \label{eqn:Maxwell-nosource}
\end{subequations}
where $\partial_t$ means $\frac{\partial}{\partial t}$, 
${\bf E}$ and ${\bf D}$ are the electric field vector and electric flux
vector, respectively, and 
${\bf H}$ and ${\bf B}$ {\bf are} the magnetic field vector and magnetic
flux vector, respectively. 
Using the dielectric constant $\varepsilon$ and the magnetic
permeability $\mu$, the ${\bf D}$ and ${\bf B}$ are expressed as: 
\begin{align}
 {\bf D}&=\varepsilon{\bf E}, &
 {\bf B}&=\mu{\bf H}.
 \label{eqn:axial-poler}
\end{align}
In the present calculation, we set $\mu=1$ in all the regions.

\begin{figure}[bt]
 \begin{center}
  \includegraphics[width=70mm]{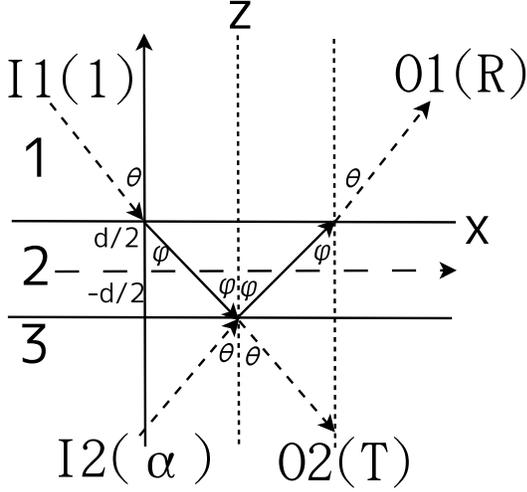}
  \caption{definition of parameters}
  \label{fig:double2}
 \end{center}
\end{figure}

In general, 
equations (\ref{eqn:Faraday}) and (\ref{eqn:no-monopole}) indicate that
there exist a scalar potential $f$ and a vector potential
${\bf A}$ which fit to
\begin{align}
 {\bf E}&=-\nabla f-\partial_t{\bf A},
 \label{eqn:gauge-field-E-general}\\
 {\bf B}&=\nabla\times{\bf A},
 \label{eqn:gauge-field-B-general}
\end{align}
and for a gauge transformation below using any scalar function $g$:
\begin{align}
 {\bf A}&\rightarrow{\bf A}+\nabla g,\\
 f&\rightarrow f-\partial_tg,
 \label{eqn:gauge-transformation}
\end{align}
${\bf E}$ and ${\bf B}$ are invariant
\cite{Jackson,Berkeley}.

~Because we are solving a reflection and refraction problem, 
we take the gauge transformation above and 
particularly we take the Lorentz gauge of:
\begin{align}
 \nabla\cdot{\bf A}&=0, &f&=0.
 \label{eqn:Lorentz-condition}
\end{align}

~Then we have expressed the incision light and the reflection light in
the region 1 as below using the vector potential:
\begin{align}
 {\bf A}^1&=\frac{-1}{i\omega}
 \begin{pmatrix}
  A^1_{TM}\cos\theta\\
  A^1_{TE}\\
  A^1_{TM}\sin\theta
 \end{pmatrix}
 e^{i\left(\omega t-n\frac{\sin\theta x-\cos\theta z}{\lambdabar}\right)},
 \label{eqn:1-1}
 \\
 {\bf A}^R&=\frac{-1}{i\omega}
 \begin{pmatrix}
  A^R_{TM}\cos\theta\\
  A^R_{TE}\\
  -A^R_{TM}\sin\theta
 \end{pmatrix}
 e^{i\left(\omega t-n\frac{\sin\theta x+\cos\theta z}{\lambdabar}\right)}.
 \label{eqn:1-R}
\end{align}
Similarly, we have expressed those in the region 2 as:
\begin{align}
 {\bf A}^{(2,1)}&=\frac{-1}{i\omega}
 \begin{pmatrix}
  A^{(2,1)}_{TM}\cos\varphi\\
  A^{(2,1)}_{TE}\\
  A^{(2,1)}_{TM}\sin\varphi
 \end{pmatrix}
 e^{i\left(\omega t-\frac{\sin\varphi x-\cos\varphi z}{\lambdabar}\right)},
 \label{eqn:2-1-refract}
 \\
 {\bf A}^{(2,2)}&=\frac{-1}{i\omega}
 \begin{pmatrix}
  A^{(2,2)}_{TM}\cos\varphi\\
  A^{(2,2)}_{TE}\\
  -A^{(2,2)}_{TM}\sin\varphi
 \end{pmatrix}
 e^{i\left(\omega t-\frac{\sin\varphi x+\cos\varphi z}{\lambdabar}\right)},
 \label{eqn:2-2-refract}
\end{align}
and we have expressed those in the region 3 as:
\begin{align}
 {\bf A}^\alpha&=\frac{-1}{i\omega}
 \begin{pmatrix}
  A^\alpha_{TM}\cos\theta\\
  A^\alpha_{TE}\\
  A^\alpha_{TM}\sin\theta
 \end{pmatrix}
 e^{i\left(\omega t-n\frac{\sin\theta x-\cos\theta z}{\lambdabar}\right)},
 \label{eqn:3-alpha}
 \\
 {\bf A}^R&=\frac{-1}{i\omega}
 \begin{pmatrix}
  A^T_{TM}\cos\theta\\
  A^T_{TE}\\
  -A^T_{TM}\sin\theta
 \end{pmatrix}
 e^{i\left(\omega t-n\frac{\sin\theta x+\cos\theta z}{\lambdabar}\right)}.
 \label{eqn:3-T}
\end{align}
Here, $\omega, \lambdabar$ and $n$ in
equations (\ref{eqn:1-1})-(\ref{eqn:3-T}) are angular frequency,
$\frac{\lambda}{2\pi}$ of light, and refractive index of region 1 and 3,
respectively.

Equations (\ref{eqn:1-1})-(\ref{eqn:3-T}) satisfy the condition of
Lorentz gauge (\ref{eqn:Lorentz-condition}), and 
variables with suffix $TM$ are for the TM mode and 
those with suffix $TE$ are for the TE mode. 

The wave equation of the vector potential:
\begin{equation}
 \frac{1}{c^2}\partial^2_t{\bf A}-\nabla^2{\bf A}=0.
  \label{eqn:wave-equation}
\end{equation}
is derived using the Maxwell's equations (\ref{eqn:Maxwell-nosource}), 
the relation between field and flux (\ref{eqn:axial-poler}), 
the relation between the vector potential ${\bf A}$ and 
the electric field ${\bf E}$ (\ref{eqn:gauge-field-E-general}), 
the relation between vector potential and magnetic flux ${\bf B}$
(\ref{eqn:gauge-field-B-general}), 
and also the relation between the light velocity $c$ and
the dielectric constant $\varepsilon$ and the magnetic 
permeability $\mu$
\begin{equation}
 c^2=\frac{1}{\varepsilon\mu},
\end{equation}
where $\mu=1$ in the present study.

Therefore, the vector potentials in each region satisfy the
dispersion relations:
\begin{align}
 \frac{\omega^2}{c_1^2}=\frac{n^2}{{\lambdabar}^2},
 &~\hspace{2em}\text{regions 1 and 3},\\
 \frac{\omega^2}{c^2}=\frac{1}{{\lambdabar}^2},
 &~\hspace{2em}\text{region 2},
\end{align}
where $c_1$ is the light velocity in regions 1 and 3, 
$c$ is the light velocity in vacuum or region 2 and 
$\lambda$ is the wave-length in vacuum.

From equations (\ref{eqn:gauge-field-E-general}), 
(\ref{eqn:gauge-field-B-general}), (\ref{eqn:1-1})-(\ref{eqn:3-T}), 
we have derived the electric field and magnetic fields in region 1 as:
\begin{align}
 {\bf E}^1(t,{\bf x})&=
 \begin{pmatrix}
  A^1_{TM}\cos\theta\\
  A^1_{TE}\\
  A^1_{TM}\sin\theta
 \end{pmatrix}
 e^{i\left(\omega t-n\frac{\sin\theta x-\cos\theta z}{\lambdabar}\right)}
 \nonumber\\
 &+
 \begin{pmatrix}
  A^R_{TM}\cos\theta\\
  A^R_{TE}\\
  -A^R_{TM}\sin\theta
 \end{pmatrix}
 e^{i\left(\omega t-n\frac{\sin\theta x+\cos\theta z}{\lambdabar}\right)},
 \label{eqn:E1}
 \\
 {\bf B}^1(t,{\bf x})&=
 \frac{n}{\omega\lambdabar}
 \begin{pmatrix}
  A^1_{TE}\cos\theta\\
  -A^1_{TM}\\
  A^1_{TE}\sin\theta
 \end{pmatrix}
 e^{i\left(\omega t-n\frac{\sin\theta x-\cos\theta z}{\lambdabar}\right)}
 \nonumber\\
 &+
 \frac{n}{\omega\lambdabar}
 \begin{pmatrix}
  -A^R_{TE}\cos\theta\\
  A^R_{TM}\\
  A^R_{TE}\sin\theta
 \end{pmatrix}
 e^{i\left(\omega t-n\frac{\sin\theta x+\cos\theta z}{\lambdabar}\right)},
 \label{eqn:B1}
\end{align}
those in region 2 as:
\begin{align}
 {\bf E}^2(t,{\bf x})&=
 \begin{pmatrix}
  A^{(2,1)}_{TM}\cos\varphi\\
  A^{(2,1)}_{TE}\\
  A^{(2,1)}_{TM}\sin\varphi
 \end{pmatrix}
 e^{i\left(\omega t-\frac{\sin\varphi x-\cos\varphi z}{\lambdabar}\right)}
 \nonumber\\
 &+
 \begin{pmatrix}
  A^{(2,2)}_{TM}\cos\varphi\\
  A^{(2,2)}_{TE}\\
  -A^{(2,2)}_{TM}\sin\varphi
 \end{pmatrix}
 e^{i\left(\omega t-\frac{\sin\varphi x+\cos\varphi z}{\lambdabar}\right)},
 \label{eqn:E2-refract}
 \\
 {\bf B}^2(t,{\bf x})&=
 \frac{1}{\omega\lambdabar}
 \begin{pmatrix}
  A^{(2,1)}_{TE}\cos\varphi\\
  -A^{(2,1)}_{TM}\\
  A^{(2,1)}_{TE}\sin\varphi
 \end{pmatrix}
 e^{i\left(\omega t-\frac{\sin\varphi x-\cos\varphi z}{\lambdabar}\right)}
 \nonumber\\
 &+
 \frac{1}{\omega\lambdabar}
 \begin{pmatrix}
  -A^{(2,2)}_{TE}\cos\varphi\\
  A^{(2,2)}_{TM}\\
  A^{(2,2)}_{TE}\sin\varphi
 \end{pmatrix}
 e^{i\left(\omega t-\frac{\sin\varphi x+\cos\varphi z}{\lambdabar}\right)},
 \label{eqn:B2-refract}
\end{align}
and also those in region 3 as:
\begin{align}
 {\bf E}^3(t,{\bf x})&=
 \begin{pmatrix}
  A^T_{TM}\cos\theta\\
  A^T_{TE}\\
  A^T_{TM}\sin\theta
 \end{pmatrix}
 e^{i\left(\omega t-n\frac{\sin\theta x-\cos\theta z}{\lambdabar}\right)}
 \nonumber\\
 &+
 \begin{pmatrix}
  A^3_{TM}\cos\theta\\
  A^3_{TE}\\
  -A^3_{TM}\sin\theta
 \end{pmatrix}
 e^{i\left(\omega t-n\frac{\sin\theta x+\cos\theta z}{\lambdabar}\right)},
 \label{eqn:E3}
 \\
 {\bf B}^3(t,{\bf x})&=
 \frac{n}{\omega\lambdabar}
 \begin{pmatrix}
  A^T_{TE}\cos\theta\\
  -A^T_{TM}\\
  A^T_{TE}\sin\theta
 \end{pmatrix}
 e^{i\left(\omega t-\frac{\sin\theta x-\cos\theta z}{\lambdabar}\right)}
 \nonumber\\
 &+
 \frac{n}{\omega\lambdabar}
 \begin{pmatrix}
  -A^3_{TE}\cos\theta\\
  A^3_{TM}\\
  A^3_{TE}\sin\theta
 \end{pmatrix}
 e^{i\left(\omega t-\frac{\sin\theta x+\cos\theta z}{\lambdabar}\right)}.
 \label{eqn:B3}
\end{align}

The ratios $\overline{R}$ and $\overline{T}$ of the output light 
intensities into regions 1 and 3 (O1 and O2) to the input
light intensity I1, respectively, can be derived from the ratios
of long time averages of z-components (vertical to the
boundary plane) of the Poynting's vectors of regions 1 and 3, 
where each of the long time average of Poynting's vector is derived as:
\begin{equation}
 \overline{\bf S}=\frac{1}{2}Re\bigl[{\bf E}\times{\bf B}^*\bigr].
 \label{eqn:PoyntingVectorDef}
\end{equation}

We have calculated the long time average of Poynting's vector of
the input light in region 1 (I1), that of output light in region 1 (O1),
that of input light in region 3 (I2) and that of output light in region
3 (O2) as following, respectively.
\begin{align}
 \overline{{\bf S}^1}&
 =n\frac{|A^1_{TM}|^2+|A^1_{TE}|^2}{2\omega\lambdabar}
 \begin{pmatrix}
  \sin\theta\\
  0\\
  -\cos\theta
 \end{pmatrix},
 \label{eqn:S^1}\\
 \overline{{\bf S}^R}&
 =n\frac{|A^R_{TM}|^2+|A^R_{TE}|^2}{2\omega\lambdabar}
 \begin{pmatrix}
  \sin\theta\\
  0\\
  \cos\theta
 \end{pmatrix},
 \label{eqn:S^R}\\
 \overline{{\bf S}^3}&
 =n\frac{|A^3_{TM}|^2+|A^3_{TE}|^2}{2\omega\lambdabar}
 \begin{pmatrix}
  \sin\theta\\
  0\\
  \cos\theta
 \end{pmatrix},
 \label{eqn:S^3}\\
 \overline{{\bf S}^T}&
 =n\frac{|A^T_{TM}|^2+|A^T_{TE}|^2}{2\omega\lambdabar}
 \begin{pmatrix}
  \sin\theta\\
  0\\
  -\cos\theta
 \end{pmatrix}.
 \label{eqn:S^T}
\end{align}

Using these equations (\ref{eqn:S^1}-\ref{eqn:S^T}), 
we have defined the light intensities $\overline{R}$ and $\overline{T}$
for TM and TE modes, respectively, as:
\begin{align}
 \overline{R}_{TM}&=\frac{|A^R_{TM}|^2}{|A^1_{TM}|^2},&
 \overline{R}_{TE}&=\frac{|A^R_{TE}|^2}{|A^1_{TE}|^2},
 \label{eqn:DefinitionOfR}\\
 \overline{T}_{TM}&=\frac{|A^T_{TM}|^2}{|A^1_{TM}|^2},&
 \overline{T}_{TE}&=\frac{|A^T_{TE}|^2}{|A^1_{TE}|^2}.
 \label{eqn:DefinitionOfT}
\end{align}
Note here that these light intensities are relative to the input
light intensity (I1) which has a value 1. 
This why symbols $\overline{R}$ and $\overline{T}$ are used instead of
$R$ and $T$.

We also have defined the ratio of an intensity of the input light in
region 3 (I2) to that of the input light in region 1 (I1) as:
\begin{equation}
 \alpha
  =\frac{|A^3_{TM}|^2}{|A^1_{TM}|^2}
  =\frac{|A^3_{TE}|^2}{|A^1_{TE}|^2}.
  \label{eqn:DeltaefOfAlpha}
\end{equation}

From equation (\ref{eqn:DeltaefOfAlpha}), it is possible to exist a
phase difference $\eta$ between two incident lights (I1) and (I2) such
as:
\begin{equation}
 \begin{pmatrix}
  A^3_{TM}\\
  A^3_{TE}
 \end{pmatrix}
 =\alpha e^{i\eta}
 \begin{pmatrix}
  A^1_{TM}\\
  A^1_{TE}
 \end{pmatrix}.
 \label{eqn:DefinitionOfAlpha}
\end{equation}
Notice that we can mathematically define modewise phase
differences to satisfy equation (\ref{eqn:DeltaefOfAlpha}), 
however, 
we are interested in a phase difference caused by an optical path
difference of a Mach-Zehnder circuit as shown in Fig.\ref{fig:m1}. 
Thus, we adapt only one phase difference $\eta$ 
which is independent of modes.

Now, ${\bf E, H, D}$ and ${\bf B}$ have to fit to the boundary 
conditions based on the Maxwell's equations 
(\ref{eqn:Maxwell-nosource}):
\begin{itemize}
 \item The parallel components to the boundaries of the 
       electric and magnetic fields $(E_x, E_y)$ and $(H_x, H_y)$
       should be continuous on the boundaries, and
 \item The vertical components to the boundaries of the 
       electric and magnetic fluxes $D_z$ and $B_z$ should be
       continuous.
\end{itemize}
From these conditions, we have derived the boundary conditions as
follows at $z=\frac{d}{2}$:
\begin{gather}
 \begin{pmatrix}
  A^1_{TM}\!\cos\theta\\
  A^1_{TE}
 \end{pmatrix}
 \!e^{i\frac{nd\cos\theta}{2\lambdabar}}
 \!\!+\!\!
 \begin{pmatrix}
  A^R_{TM}\!\cos\theta\\
  A^R_{TE}
 \end{pmatrix}
 \!e^{-i\frac{nd\cos\theta}{2\lambdabar}}
 \nonumber\\
 =\!
 \begin{pmatrix}
  A^{(2,1)}_{TM}\!\cos\varphi\\
  A^{(2,1)}_{TE}
 \end{pmatrix}
 \!e^{i\frac{d\cos\varphi}{2\lambdabar}}
 \!\!+\!\!
 \begin{pmatrix}
  A^{(2,2)}_{TM}\!\cos\varphi\\
  A^{(2,2)}_{TE}
 \end{pmatrix}\!
 e^{-i\frac{d\cos\varphi}{2\lambdabar}},
 \label{eqn:+d/2-E}\\
 \varepsilon_1\sin\theta
 \bigl(
 A^1_{TM}e^{i\frac{nd\cos\theta}{2\lambdabar}}
 -
 A^R_{TM}e^{-i\frac{nd\cos\theta}{2\lambdabar}}
 \bigr)
 \nonumber\\
 =
 \varepsilon_2\sin\varphi
 \bigl(
 A^{(2,1)}_{TM}e^{i\frac{nd\cos\varphi}{2\lambdabar}}
 -
 A^{(2,2)}_{TM}e^{-i\frac{nd\cos\varphi}{2\lambdabar}}
 \bigr),
 \label{eqn:+d/2-D}\\
 \frac{n}{\omega\lambdabar}\!\!
 \left[\!\!
 \begin{pmatrix}
  \!\!A^1_{TE}\!\cos\theta\!\!\\
  -A^1_{TM}
 \end{pmatrix}
 e^{i\frac{nd\cos\theta}{2\lambdabar}}
 \!\!+\!\!
 \begin{pmatrix}
  \!\!-A^R_{TE}\!\cos\theta\!\!\\
  A^R_{TM}
 \end{pmatrix}
 e^{-i\frac{nd\cos\theta}{2\lambdabar}}\!\!
 \right]
 \nonumber\\
 ~\hspace{-1em}
 =\!
 \frac{1}{\omega\lambdabar}\!\!
 \left[\!\!
 \begin{pmatrix}
  \!\!A^{(\!2\!,\!1\!)}_{TE}\!\cos\varphi\!\!\\
  -A^{(\!2\!,\!1\!)}_{TM}
 \end{pmatrix}\!\!
 e^{i\frac{nd\cos\varphi}{2\lambdabar}}
 \!\!+\!\!
 \begin{pmatrix}
  \!\!-A^{(\!2\!,\!2\!)}_{TE}\!\cos\varphi\!\!\\
  A^{(\!2\!,\!2\!)}_{TM}
 \end{pmatrix}\!\!
 e^{-i\frac{nd\cos\varphi}{2\lambdabar}}\!\!
 \right],
 \label{eqn:+d/2-H}\\
 \frac{n\sin\theta}{\omega\lambdabar}
 \bigl(
 A^1_{TE}e^{i\frac{nd\cos\theta}{2\lambdabar}}
 +
 A^R_{TE}e^{-i\frac{nd\cos\theta}{2\lambdabar}}
 \bigr)
 \nonumber\\
 =
 \frac{\sin\varphi}{\omega\lambdabar}
 \bigl(
  A^{(2,1)}_{TE}e^{i\frac{nd\cos\varphi}{2\lambdabar}}
 +
 A^{(2,2)}_{TE}e^{-i\frac{nd\cos\varphi}{2\lambdabar}}
 \bigr).
 \label{eqn:+d/2-B}
\end{gather}

These equations (\ref{eqn:+d/2-E})-(\ref{eqn:+d/2-B}) express the boundary
conditions for $E_x$ and $E_y$, that for $D_z$, those for $H_x$ and $H_y$
and the boundary condition for $B_z$, respectively. 
Because the equations (\ref{eqn:+d/2-D}) and (\ref{eqn:+d/2-B}) are
included in equations (\ref{eqn:+d/2-E}) and (\ref{eqn:+d/2-H}), 
we have selected the independent boundary condition equations as:
\begin{gather}
 \cos\theta\bigl(
 A^1_{TM}e^{i\frac{nd\cos\theta}{2\lambdabar}}
 +
 A^R_{TM}e^{-i\frac{nd\cos\theta}{2\lambdabar}}
 \bigr)
 \nonumber\\
 =
 \cos\varphi\bigl(
 A^{(2,1)}_{TM}e^{i\frac{d\cos\varphi}{2\lambdabar}}
 +
 A^{(2,2)}_{TM}e^{-i\frac{d\cos\varphi}{2\lambdabar}}
 \bigr),
 \label{eqn:+d/2-TM+}\\[1ex]
 n\bigl(
 A^1_{TM}e^{i\frac{nd\cos\theta}{2\lambdabar}}
 -
 A^R_{TM}e^{-i\frac{nd\cos\theta}{2\lambdabar}}
 \bigr)
 \nonumber\\
 =
 A^{(2,1)}_{TM}e^{i\frac{d\cos\varphi}{2\lambdabar}}
 -
 A^{(2,2)}_{TM}e^{-i\frac{d\cos\varphi}{2\lambdabar}},
 \label{eqn:+d/2-TM-}\\[1ex]
 A^1_{TE}e^{i\frac{nd\cos\theta}{2\lambdabar}}
 +
 A^R_{TE}e^{-i\frac{nd\cos\theta}{2\lambdabar}}
 \nonumber\\
 =
 A^{(2,1)}_{TE}e^{i\frac{d\cos\varphi}{2\lambdabar}}
 +
 A^{(2,2)}_{TE}e^{-i\frac{d\cos\varphi}{2\lambdabar}},
 \label{eqn:+d/2-TE+}\\[1ex]
 n\cos\theta\bigl(
 A^1_{TE}e^{i\frac{nd\cos\theta}{2\lambdabar}}
 -
 A^R_{TE}e^{-i\frac{nd\cos\theta}{2\lambdabar}}
 \bigr)
 \nonumber\\
 =
 \cos\varphi\bigl(
 A^{(2,1)}_{TE}e^{i\frac{d\cos\varphi}{2\lambdabar}}
 -
 A^{(2,2)}_{TE}e^{-i\frac{d\cos\varphi}{2\lambdabar}}
 \bigr).
 \label{eqn:+d/2-TE-}
\end{gather}
These are two sets of boundary condition equations for TM and TE mode,
respectively.

We have found from equations (\ref{eqn:+d/2-TM+}) and
(\ref{eqn:+d/2-TM-}) that there exist two $2\times 2$ matrices
$M_{TM_+}$ and $N_{TM_+}$ such as:
\begin{equation}
 M_{TM_+}
  \begin{pmatrix}
   A^1_{TM}\\
   A^R_{TM}
  \end{pmatrix}
  =
  N_{TM_+}
  \begin{pmatrix}
   A^{(2,1)}_{TM}\\
   A^{(2,2)}_{TM}
  \end{pmatrix},
\end{equation}
and we have solved the equation as:
\begin{align}
 \begin{pmatrix}
  A^{(2,1)}_{TM}\\
  A^{(2,2)}_{TM}
 \end{pmatrix}
 &=N^{-1}_{TM_+}M_{TM_+}
 \begin{pmatrix}
  A^1_{TM}\\
  A^R_{TM}
 \end{pmatrix},\\
 &=P_{TM_+}
 \begin{pmatrix}
  A^1_{TM}\\
  A^R_{TM}
 \end{pmatrix},
 \label{eqn:+d/2-TM-solution}
\end{align}
where
\begin{equation}
 P_{TM_+} =N^{-1}_{TM_+}M_{TM_+}.
\end{equation}

We have also derived another $2\times 2$ matrices for TE mode from
(\ref{eqn:+d/2-TE+}) and (\ref{eqn:+d/2-TE-}), and the solution is:
\begin{equation}
 \begin{pmatrix}
  A^{(2,1)}_{TE}\\
  A^{(2,2)}_{TE}
 \end{pmatrix}
 =P_{TE_+}
 \begin{pmatrix}
  A^1_{TE}\\
  A^R_{TE}
 \end{pmatrix}.
 \label{eqn:+d/2-TE-solution}
\end{equation}

Similarly, we have derived the boundary conditions at $z=-\frac{d}{2}$
as:
\begin{gather}
 \begin{pmatrix}
  \!A^{(\!2\!,\!1\!)}_{TM}\cos\varphi\!\\
  A^{(\!2\!,\!1)}_{TE}
 \end{pmatrix}
 e^{-i\frac{d\cos\varphi}{2\lambdabar}}
 +
 \begin{pmatrix}
  \!A^{(\!2\!,\!2\!)}_{TM}\cos\varphi\!\\
  A^{(\!2\!,\!2\!)}_{TE}
 \end{pmatrix}
 e^{i\frac{d\cos\varphi}{2\lambdabar}}
 \nonumber\\
 =
 \begin{pmatrix}
  \!A^T_{TM}\!\cos\theta\!\\
  A^T_{TE}
 \end{pmatrix}
 e^{-i\frac{nd\cos\theta}{2\lambdabar}}
 +
 \begin{pmatrix}
  \!A^3_{TM}\cos\theta\!\\
  A^3_{TE}
 \end{pmatrix}
 e^{i\frac{nd\cos\theta}{2\lambdabar}},
 \\
 \varepsilon_2\sin\varphi
 \bigl(
 A^{(2,1)}_{TM}e^{-i\frac{d\cos\varphi}{2\lambdabar}}
 +
 A^{(2,2)}_{TM}e^{i\frac{d\cos\varphi}{2\lambdabar}}
 \bigr)
 \nonumber\\
 =
 \varepsilon_1\sin\theta
 \bigl(
 A^T_{TM}e^{-i\frac{nd\cos\theta}{2\lambdabar}}
 -
 A^3_{TM}e^{i\frac{nd\cos\theta}{2\lambdabar}}
 \bigr),\\
 \frac{1}{\omega\lambdabar}\!\!
 \left[\!\!
 \begin{pmatrix}
  \!A^{(\!2\!,\!1\!)}_{TE}\cos\varphi\!\\
  -A^{(\!2\!,\!1\!)}_{TM}
 \end{pmatrix}
 \!e^{-i\frac{nd\cos\varphi}{2\lambdabar}}
 \!\!\!+\!\!
 \begin{pmatrix}
  \!-\!A^{(\!2\!,\!2\!)}_{TE}\cos\varphi\!\\
  A^{(\!2\!,\!2\!)}_{TM}
 \end{pmatrix}
 e^{i\frac{d\cos\varphi}{2\lambdabar}}
 \right]
 \nonumber\\
 =\!\!
 \frac{n}{\omega\lambdabar}\!\!
 \left[\!\!
 \begin{pmatrix}
  \!A^T_{TE}\cos\theta\!\\
  -A^T_{TM}
 \end{pmatrix}\!\!
 e^{-i\frac{nd\cos\theta}{2\lambdabar}}
 \!\!+\!\!
 \begin{pmatrix}
  \!-A^3_{TE}\cos\theta\!\\
  A^3_{TM}
 \end{pmatrix}\!\!
 e^{i\frac{nd\cos\theta}{2\lambdabar}}\!\!
 \right],\\
 \frac{\sin\varphi}{\omega\lambdabar}
 \bigl(
 A^{(2,1)}_{TE}e^{-i\frac{d\cos\varphi}{2\lambdabar}}
 +
 A^{(2,2)}_{TE}e^{i\frac{d\cos\varphi}{2\lambdabar}}
 \bigr)
 \nonumber\\
 =
 \frac{n\sin\theta}{\omega\lambdabar}
 \bigl(
 A^T_{TE}e^{-i\frac{nd\cos\theta}{2\lambdabar}}
 +
 A^3_{TE}e^{i\frac{nd\cos\theta}{2\lambdabar}}
 \bigr),
\end{gather}
and we have found the independent boundary conditions as:
\begin{gather}
 \cos\varphi\bigl(
 A^{(2,1)}_{TM}e^{-i\frac{d\cos\varphi}{2\lambdabar}}
 +
 A^{(2,2)}_{TM}e^{i\frac{d\cos\varphi}{2\lambdabar}}
 \bigr)
 \nonumber\\
 =
 \cos\theta\bigl(
 A^T_{TM}e^{-i\frac{nd\cos\theta}{2\lambdabar}}
 +
 A^3_{TM}e^{i\frac{nd\cos\theta}{2\lambdabar}}
 \bigr),
 \label{eqn:-d/2-TM+}\\[1ex]
 \bigl(
 A^{(2,1)}_{TM}e^{-i\frac{d\cos\varphi}{2\lambdabar}}
 -
 A^{(2,2)}_{TM}e^{i\frac{d\cos\varphi}{2\lambdabar}}
 \bigr)
 \nonumber\\
 =
 n\bigl(
 A^T_{TM}e^{-i\frac{nd\cos\theta}{2\lambdabar}}
 -
 A^3_{TM}e^{i\frac{d\cos\theta}{2\lambdabar}}
 \bigr),
 \label{eqn:-d/2-TM-}\\[1ex]
 A^{(2,1)}_{TE}e^{-i\frac{d\cos\varphi}{2\lambdabar}}
 +
 A^{(2,2)}_{TE}e^{i\frac{d\cos\varphi}{2\lambdabar}}
 \nonumber\\
 =
 A^T_{TE}e^{-i\frac{nd\cos\theta}{2\lambdabar}}
 +
 A^3_{TE}e^{i\frac{nd\cos\theta}{2\lambdabar}},
 \label{eqn:-d/2-TE+}\\[1ex]
 \cos\varphi\bigl(
 A^{(2,1)}_{TE}e^{-i\frac{d\cos\varphi}{2\lambdabar}}
 -
 A^{(2,2)}_{TE}e^{i\frac{d\cos\varphi}{2\lambdabar}}
 \bigr)
 \nonumber\\
 =
 n\cos\theta\bigl(
 A^T_{TE}e^{-i\frac{nd\cos\theta}{2\lambdabar}}
 -
 A^3_{TE}e^{i\frac{nd\cos\theta}{2\lambdabar}}
 \bigr).
 \label{eqn:-d/2-TE-}
\end{gather}

We have derived a $2\times 2$ matrices $P_{TM_-}$ and $P_{TE_-}$ from
equations (\ref{eqn:-d/2-TM+})-(\ref{eqn:-d/2-TE-}) as:
\begin{align}
 \begin{pmatrix}
  A^{(2,1)}_{TM}\\
  A^{(2,2)}_{TM}
 \end{pmatrix}
 &=P_{TM_-}
 \begin{pmatrix}
  A^3_{TM}\\
  A^T_{TM}
 \end{pmatrix},
 \label{eqn:-d/2-TM-solution}\\
 \begin{pmatrix}
  A^{(2,1)}_{TE}\\
  A^{(2,2)}_{TE}
 \end{pmatrix}
 &=P_{TE_-}
 \begin{pmatrix}
  A^3_{TE}\\
  A^T_{TE}
 \end{pmatrix}.
 \label{eqn:-d/2-TE-solution}
\end{align}

Combining those with equations (\ref{eqn:+d/2-TM-solution}) and
(\ref{eqn:-d/2-TM-solution}), 
we have calculated 
the dependences of the output lights $A^R_{TM}$ and $A^T_{TM}$ on
$A^1_{TM}$ and $A^3_{TM}$ for TM mode as:
\begin{align}
  A^R_{TM}&=-\frac
 {\begin{bmatrix}
   \!\!
   \bigl(\!\cos^2\!\theta\!-\!n^2\!\cos^2\!\varphi\!\bigr)
   \sin\bigl(\frac{d\cos\varphi}{\lambdabar}\bigr)A^1_{TM}\!\!\\
   +
   i2n\cos\theta\cos\varphi
   A^3_{TM}
  \end{bmatrix}}
 {\begin{bmatrix}
   \bigl(\!\cos^2\theta\!+\!n^2\cos^2\varphi\!\bigr)\!
   \sin\bigl(\frac{d\cos\varphi}{\lambdabar}\bigr)\\
   \!-\!
   i2n\cos\theta\cos\varphi
   \cos\bigl(\frac{d\cos\varphi}{\lambdabar}\bigr)\!
  \end{bmatrix}
 }
 e^{i\!\frac{nd\!\cos\!\theta}{\lambdabar}},
 \label{eqn:A^R_TM}\\
 A^T_{TM}&=-\frac
 {\begin{bmatrix}
   in\cos\theta\cos\varphi A^1_{TM}+\\
   \!\!\bigl(\!\cos^2\theta\!-\!n^2\!\cos^2\varphi\!\bigr)\!
   \sin\bigl(\frac{d\cos\varphi}{\lambdabar}\bigr)A^3_{TM}\!\!
  \end{bmatrix}}
 {\begin{bmatrix}
   \bigl(\!\cos^2\theta\!+\!n^2\!\cos^2\varphi\!\bigr)\!
   \sin\bigl(\frac{d\cos\varphi}{\lambdabar}\bigr)\\
   -
   i2n\cos\theta\cos\varphi
   \cos\bigl(\frac{d\cos\varphi}{\lambdabar}\bigr)
  \end{bmatrix}}
   e^{i\frac{nd\cos\theta}{2\lambdabar}}.
 \label{eqn:A^T_TM}
\end{align}

We have also calculated those for TE mode as:
\begin{align}
  A^R_{TE}&=
  -\frac
  {\begin{bmatrix}
    \!\!\bigl(\!\cos^2\!\varphi\!-\!n^2\cos^2\!\theta\!\bigr)\!
    \sin\bigl(\frac{d\cos\varphi}{\lambdabar}\bigr)A^1_{TE}\!\!\\
    +
    i2n\cos\theta\cos\varphi
    A^3_{TE}
   \end{bmatrix}}
  {\begin{bmatrix}
    \bigl(\cos^2\!\varphi\!+\!n^2\!\cos^2\!\theta\!\bigr)\!
    \sin\bigl(\frac{d\cos\varphi}{\lambdabar}\bigr)\\
    -
    i2n\cos\theta\cos\varphi
    \cos\bigl(\frac{d\cos\varphi}{\lambdabar}\bigr)
   \end{bmatrix}}
 e^{i\frac{nd\cos\theta}{\lambdabar}},
 \label{eqn:A^R_TE}\\
 A^T_{TE}&=
  -\frac
  {\begin{bmatrix}
    i2n\cos\theta\cos\varphi A^1_{TE}
    +\\
    \!\!\bigl(\cos^2\!\varphi\!-\!n^1\!\cos\!\theta\!\bigr)\!
    \sin\bigl(\frac{d\cos\varphi}{\lambdabar}\bigr)
    A^3_{TE}\!\!
   \end{bmatrix}}
 {\begin{bmatrix}
   \bigl(\cos^2\!\varphi\!+\!n^2\!\cos^2\!\theta\!\bigr)\!
   \sin\bigl(\frac{d\cos\varphi}{\lambdabar}\bigr)\\
   -
   i2n\cos\theta\cos\varphi
   \cos\bigl(\frac{d\cos\varphi}{\lambdabar}\bigr)
  \end{bmatrix}}
 e^{i\frac{d\cos\theta}{\lambdabar}}.
 \label{eqn:A^T_TE}
\end{align}

We have combined 
equations (\ref{eqn:DefinitionOfR}), (\ref{eqn:DefinitionOfAlpha}), 
(\ref{eqn:A^R_TM}) and (\ref{eqn:A^R_TE}) to derive:
\begin{align}
\overline{R}_{TM}&=
 \frac
 {\begin{bmatrix}
   (\cos^2\!\theta\!-\!n^2\!\cos^2\!\varphi)^2\!
   \sin^2\bigl(\frac{d\cos\varphi}{\lambdabar}\bigr)\\
   +4\alpha n^2\cos^2\!\theta\cos^2\!\varphi\\
   -\sqrt{\alpha}n\sin\!\eta\cos\!\theta\cos\!\varphi\\
   \times
   (\cos^2\!\theta\!-\!n\!\cos^2\!\varphi)\!
   \sin\bigl(\frac{d\cos\varphi}{\lambdabar}\bigr)
  \end{bmatrix}}
 {\begin{bmatrix}
   4n^2\cos^2\!\theta\cos^2\!\varphi+\\
   (\cos^2\!\theta\!-\!n^2\cos^2\!\varphi)^2\!
   \sin^2\bigl(\frac{d\cos\varphi}{\lambdabar}\bigr)
  \end{bmatrix}}
 ,\label{eqn:R_TM-refract}\\
 \overline{R}_{TE}&=
 \frac
 {\begin{bmatrix}
   (\cos^2\!\varphi\!-\!n^2\!\cos^2\!\theta)^2\!
   \sin^2\bigl(\frac{d\cos\varphi}{\lambdabar}\bigr)\\
   +\!4\alpha n^2\cos^2\theta\cos^2\varphi\\
   -4n\sqrt{\alpha}\sin\!\eta\cos\!\theta\cos\!\varphi\\
   \times
   (\cos^2\!\varphi\!-\!n^2\!\cos^2\!\theta)\!
   \sin\bigl(\frac{d\cos\varphi}{\lambdabar}\bigr)
  \end{bmatrix}}
 {\begin{bmatrix}
   4n^2\cos^2\!\theta\cos^2\!\varphi+\\
   (\cos^2\!\varphi\!-\!n^2\!\cos^2\!\theta)^2\!
   \sin^2\bigl(\frac{d\cos\varphi}{\lambdabar}\bigr)
  \end{bmatrix}}
   .\label{eqn:R_TE-refract}
\end{align}

We can also calculate $\overline{T}_{TM}$ and $\overline{T}_{TE}$ which
are not shown, and easily check the relations:
\begin{gather}
 \overline{R}_{TM}+\overline{T}_{TM}=1+\alpha,\\
 \overline{R}_{TE}+\overline{T}_{TE}=1+\alpha,
\end{gather}
which tell that the energy in this model is conserved.

Furthermore, from Snell's equation:
\begin{equation}
 \sin\varphi=n\sin\theta,
  \label{eqn:Snell'seq}
\end{equation}
and a definition of variable $\kappa$:
\begin{equation}
\kappa=\tan^2\theta,
 \label{eqn:theta-kappa}
\end{equation}
we have led the relations among $\kappa, \theta, n$ and $\varphi$
expressed as:
\begin{gather}
 \frac{\cos\varphi}{\cos\theta}=\sqrt{1-(n^2-1)\kappa},
 \label{eqn:cosvarphiovercostheta-kappa}\\
 \cos\varphi=\sqrt{\frac{1-(n^2-1)\kappa}{1+\kappa}},
 \label{eqn:cosvarphi-kappa}
\end{gather}
Using the relations (\ref{eqn:cosvarphiovercostheta-kappa}) and 
(\ref{eqn:cosvarphi-kappa}), 
we have finally derived $\overline{R}_{TM}$ and $\overline{R}_{TE}$ as
functions of $d, \lambdabar, \alpha, n$, and $\kappa$ as:
\begin{align}
 \overline{R}_{TM}&=
 \frac
 {\begin{bmatrix}
   \!(n^2\!-\!1)^2\!(1\!-n^2\!\kappa)^2\!
   \sin^2\bigl(\!\frac{d}{\lambdabar}\!
   \sqrt{\frac{1\!-\!(n^2\!-\!1)\kappa}{1\!+\!\kappa}}\bigr)\!\\
   +\!4\alpha n^2(1-(n^2-1)\kappa)\\
   +4\!\sqrt{\!\alpha\!}\sin\!\eta n (n^2\!-\!1)\\
   \times\!\sqrt{1\!-\!(n^2\!-\!1)\kappa}
   \sin\!\bigl(\!\frac{d}{\lambdabar}
   \sqrt{\!\frac{1\!-\!(n^2\!-\!1)\kappa}{1+\kappa}}\bigr)
  \end{bmatrix}}
 {\begin{bmatrix}
   (n^2\!-\!1)^2\!(1\!-\!n^2\kappa)^2
   \sin^2\!\bigl(\!\frac{d}{\lambdabar}\!
   \sqrt{\!\frac{1\!-\!(n^2\!-\!1)\kappa}{1+\kappa}}\bigr)\\
   +\!4n^2\!(1\!-\!(n^2\!-\!1)\kappa) 
  \end{bmatrix}}
 ,\label{eqn:R_TM-refract-kappa}\\
 \overline{R}_{TE}&=
 \frac
 {\begin{bmatrix}
   (n^1\!-\!1)^2\!(1\!+\!\kappa)^2
   \sin^2\!\bigl(\!\frac{d}{\lambdabar}\!
   \sqrt{\!\frac{1\!-\!(n^2\!-\!1)\kappa}{1+\kappa}}\bigr)\\
   +4\alpha n^2(1\!-\!(n^2\!-\!1)\kappa)\\
   +4\sqrt{\alpha}\sin\!\eta n\!(n^2\!-\!1)(1\!+\!\kappa)\\
   \times\!\sqrt{1\!-\!(n^2\!-\!1)\kappa}
   \sin\!\bigl(\!\frac{d}{\lambdabar}\!
   \sqrt{\!\frac{1\!-\!(n^2\!-\!1)x}{1+\kappa}}\bigr)
  \end{bmatrix}}
 {\begin{bmatrix}
   (n^1\!-\!1)^2(1\!+\!\kappa)^2
   \sin\!\bigl(\!\frac{d}{\lambdabar}\!
   \sqrt{\!\frac{1\!-\!(n^2\!-\!1)\kappa}{1+\kappa}}\bigr)\\
   +4n^2(1\!-\!(n^2\!-\!1)\kappa)
  \end{bmatrix}}
 .\label{eqn:R_TE-refract-kappa}
\end{align}

\section{Output intensity when the injection angle exceeds the
 critical angle}\label{sec:evanescent}
When the injection angle exceeds the critical angle, 
the vector potential in region 2 is expressed by, 
instead of equation (\ref{eqn:2-1-refract}) and (\ref{eqn:2-2-refract}),
a linear combination of:
\begin{align}
 {\bf A}^{(2,1)}&=\frac{-1}{i\omega}\!\!
 \begin{pmatrix}
  A^{(2,1)}_{TM}\sinh\varphi'\\
  A^{(2,1)}_{TE}\\
  iA^{(2,1)}_{TM}\cosh\varphi'
 \end{pmatrix}\!\!
 e^{i\left(\omega t
 \!-\!\frac{\cosh\!\varphi'\!x\!-\!i\!\sinh\!\varphi'\!z}{\lambdabar}\right)},
 \label{eqn:2-1-eva}
 \\
 {\bf A}^{(2,2)}&=\frac{-1}{i\omega}\!\!
 \begin{pmatrix}
  A^{(2,2)}_{TM}\sinh\varphi'\\
  A^{(2,2)}_{TE}\\
  -iA^{(2,2)}_{TM}\cosh\varphi'
 \end{pmatrix}\!\!
 e^{i\left(\omega t\!-\!
 \frac{\cosh\!\varphi'\!x\!+\!i\!\sinh\!\varphi'\!z}{\lambdabar}\right)}.
 \label{eqn:2-2-eva}
\end{align}

\begin{figure}[bt]
 \begin{center}
  \includegraphics[width=70mm]{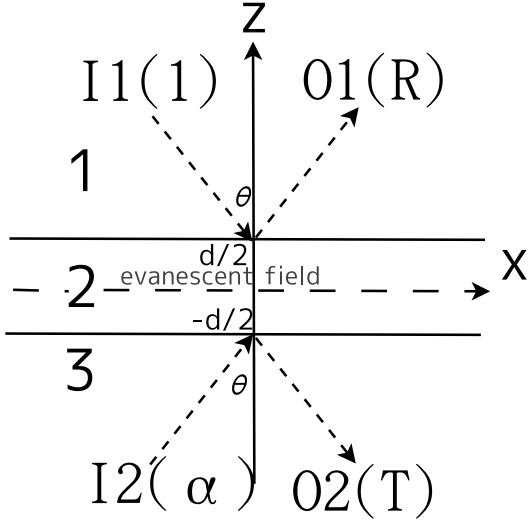}
  \caption{definition of parameters}
  \label{fig:double-eva}
 \end{center}
\end{figure}

With the boundary between region 1 and region 2 and 
in the large $d$ limit,
which means to remove region 3, 
the injected light is reflected back to region 1 by the 'total
reflection', 
and there exists no light propagating into region 2 to $z$ direction. 
Instead, the evanescent light expressed by equation (\ref{eqn:2-2-eva})
is generated which decays in a short distance of order of wavelength
exponentially according to distance from the boundary. 
Both (\ref{eqn:2-1-eva}) and (\ref{eqn:2-2-eva}) are solutions of 
the Maxwell's equations, 
but usually (\ref{eqn:2-1-eva}) is not considered 
because the intensity of the electro-magnetic field becomes infinity at
$z=-\infty$, 
contrary to locality that is common understanding of physics. 

However, in the present model where the region 2 is not infinite 
but has a definite value of width (thickness), 
we have to be careful that equation (\ref{eqn:2-1-eva}) is finite
intensity everywhere in region 2 and therefore we cannot neglect
(\ref{eqn:2-1-eva}).

Different from the usual light propagation expressed by
equations (\ref{eqn:2-1-refract}) and (\ref{eqn:2-2-refract}), 
the energy propagation in region 2 appears only in the cross term of
(\ref{eqn:2-1-eva}) and (\ref{eqn:2-2-eva}) expressing the interaction
of evanescent lights which decays exponentially to the directions
of $+z$ and $-z$, respectively. 
This is supported with the fact that, in the large $d$ limit
with fixed boundary between region 1 and region 2, 
where region 3 is removed, 
there occurs the total reflection and (\ref{eqn:2-2-eva}) becomes the
only solution for region 2, 
and there exists no energy propagation to $-z$ direction in region 2.

We have calculated the electric field and magnetic field in region 2
from equations (\ref{eqn:2-1-eva}) and (\ref{eqn:2-2-eva}) as:
\begin{align}
 {\bf E}^2&=
 \begin{pmatrix}
  A^{(2,1)}_{TM}\sinh\varphi'\\
  A^{(2,1)}_{TE}\\
  iA^{(2,1)}_{TM}\cosh\varphi'
 \end{pmatrix}
 e^{i\left(\omega t-\frac{\cosh\varphi'x+i\sinh\varphi'z}{\lambdabar}\right)}
 \nonumber\\
 &+
 \begin{pmatrix}
  A^{(2,2)}_{TM}\sinh\varphi'\\
  A^{(2,2)}_{TE}\\
  -iA^{(2,2)}_{TM}\cosh\varphi'
 \end{pmatrix}
 e^{i\left(\omega t-\frac{\cosh\varphi'x-i\sinh\varphi'z}{\lambdabar}\right)},
 \label{eqn:E2-eva}\\
 {\bf B}^2&=\frac{1}{\omega\lambdabar}
 \begin{pmatrix}
  -iA^{(2,1)}_{TE}\sinh\varphi'\\
  -iA^{(2,1)}_{TM}\\
  A^{(2,1)}_{TE}\cosh\varphi'
 \end{pmatrix}
 e^{i\left(\omega t-\frac{\cosh\varphi'x+i\sinh\varphi'z}{\lambdabar}\right)}
 \nonumber\\
 &+\frac{1}{\omega\lambdabar}
 \begin{pmatrix}
  iA^{(2,2)}_{TE}\sinh\varphi'\\
  iA^{(2,2)}_{TM}\\
  A^{(2,2)}_{TE}\cosh\varphi'
 \end{pmatrix}
 e^{i\left(\omega t-\frac{\cosh\varphi'x-i\sinh\varphi'z}{\lambdabar}\right)}.
 \label{eqn:B2-eva}
\end{align}

With derivation similar to that in section \ref{sec:refraction}, 
and using equations (\ref{eqn:E1}, \ref{eqn:E2-eva}, \ref{eqn:E3}, 
\ref{eqn:B1}, \ref{eqn:B2-eva}, \ref{eqn:B3}), 
we have derived 
the independent boundary conditions at $z=\frac{d}{2}$ as:
\begin{gather}
 \cos\theta
 (A^1_{TM}e^{i\frac{nd\cos\theta}{2\lambdabar}}
 +A^R_{TM}e^{-i\frac{nd\cos\theta}{2\lambdabar}})
 \nonumber\\
 =
 \sinh\varphi'
 (A^{(2,1)}_{TM}e^{\frac{d\sinh\varphi'}{2\lambdabar}}
 +
 A^{(2,2)}_{TM}e^{-\frac{d\sinh\varphi'}{2\lambdabar}}),
 \label{eqn:+d/2TM+eva}\\[1ex]
 n
 (A^1_{TM}e^{i\frac{nd\cos\theta}{2\lambdabar}}
 -
 A^R_{TM}e^{-i\frac{nd\cos\theta}{2\lambdabar}})
 \nonumber\\
 =
 i
 (A^{(2,1)}_{TM}e^{\frac{d\sinh\varphi'}{2\lambdabar}}
 -A^{(2,2)}_{TM}e^{-\frac{d\sinh\varphi'}{2\lambdabar}}),
 \label{eqn:+d/2TM-eva}\\[1ex]
 A^1_{TE}e^{i\frac{nd\cos\theta}{2\lambdabar}}
 +
 A^R_{TE}e^{-i\frac{nd\cos\theta}{2\lambdabar}}
 \nonumber\\
 =
 A^{(2,1)}_{TE}e^{\frac{d\sinh\varphi'}{2\lambdabar}}
 +
 A^{(2,2)}_{TE}e^{-\frac{d\sinh\varphi'}{2\lambdabar}},
 \label{eqn:+d/2TE+eva}\\[1ex]
 n\cos\theta
 (A^1_{TE}e^{i\frac{nd\cos\theta}{2\lambdabar}}
 -
 A^R_{TE}e^{-i\frac{nd\cos\theta}{2\lambdabar}})
 \nonumber\\
 =
 -i\sinh\varphi'
 (A^{(2,1)}_{TE}e^{\frac{d\sinh\varphi'}{2\lambdabar}}
 -
 A^{(2,2)}_{TE}e^{-\frac{d\sinh\varphi'}{2\lambdabar}}).
 \label{eqn:+d/2TE-eva}
\end{gather}
and
\begin{gather}
 \begin{pmatrix}
   A^{(2,1)}_{TM}e^{\frac{d\sinh\varphi'}{2\lambdabar}}\\
  A^{(2,2)}_{TM}e^{-\frac{d\sinh\varphi'}{2\lambdabar}}
 \end{pmatrix}
 \nonumber\\
 =\!\!
 \begin{pmatrix}
  \frac{\cos\!\theta\!+\!in\!\sinh\!\varphi'}{2\sinh\varphi'}&
  \frac{\cos\!\theta\!-\!in\!\sinh\!\varphi'}{2\sinh\varphi'}\\
  \frac{\cos\!\theta\!-\!in\!\sinh\!\varphi'}{2\sinh\varphi'}&
  \frac{\cos\!\theta\!+\!in\!\sinh\!\varphi'}{2\sinh\varphi'}
 \end{pmatrix}\!\!
 \begin{pmatrix}
  A^3_{TM}e^{i\frac{d\cos\theta}{2\lambdabar}}\\
  A^T_{TM}e^{-i\frac{d\cos\theta}{2\lambdabar}}
 \end{pmatrix},
 \label{eqn:A21+d/2}\\
 \begin{pmatrix}
  A^{(2,1)}_{TE}e^{\frac{d\omega\sinh\varphi'}{2c_2}}\\
  A^{(2,2)}_{TE}e^{-\frac{d\omega\sinh\varphi'}{2c_2}}
 \end{pmatrix}
 \nonumber\\
 =\!\!
 \begin{pmatrix}
 \!\! \frac{n\cos\!\theta\!+\!i\sinh\!\varphi'}{i\sinh\varphi'}&
 \!\!-\frac{n\cos\!\theta\!-\!i\sinh\!\varphi'}{i\sinh\varphi'}\\
 \!\!-\frac{n\cos\!\theta\!-\!i\sinh\!\varphi'}{i\sinh\varphi'}&
 \!\! \frac{n\cos\!\theta\!+\!i\sinh\!\varphi'}{i\sinh\varphi'}
 \end{pmatrix}\!\!
 \begin{pmatrix}
   A^3_{TE}e^{i\frac{nd\cos\!\theta}{2\lambdabar}}\\
  A^T_{TE}e^{-i\frac{nd\cos\!\theta}{2\lambdabar}}
 \end{pmatrix}.
 \label{eqn:A22+d/2}
\end{gather}

We have also derived those at $z=-\frac{d}{2}$ as:
\begin{gather}
 \sinh\varphi'
 (A^{(2,1)}_{TM}e^{-\frac{d\sinh\varphi'}{2\lambdabar}}
 +
 A^{(2,2)}_{TM}e^{\frac{d\sinh\varphi'}{2\lambdabar}})
 \nonumber\\
 =\cos\theta
 (A^3_{TM}e^{i\frac{nd\cos\theta}{2\lambdabar}}
 +A^T_{TM}e^{-i\frac{nd\cos\theta}{2\lambdabar}}),
 \label{eqn:-d/2TM+eva}\\[1ex]
 i
 (A^{(2,1)}_{TM}e^{-\frac{d\sinh\varphi'}{2\lambdabar}}
 -
 A^{(2,2)}_{TM}e^{\frac{d\sinh\varphi'}{2\lambdabar}})
 \nonumber\\
 =
 -n
 (A^3_{TM}e^{i\frac{nd\cos\theta}{2\lambdabar}}
 -
 A^T_{TM}e^{-i\frac{nd\cos\theta}{2\lambdabar}}),
 \label{eqn:-d/2TM-eva}\\[1ex]
 A^{(2,1)}_{TE}e^{-\frac{d\sinh\varphi'}{2\lambdabar}}
 +
 A^{(2,2)}_{TE}e^{\frac{d\sinh\varphi'}{2\lambdabar}}
 \nonumber\\
 =
 A^3_{TE}e^{i\frac{nd\cos\theta}{2\lambdabar}}
 +
 A^T_{TE}e^{-i\frac{nd\cos\theta}{2\lambdabar}},
 \label{eqn:-d/2TE+eva}\\[1ex]
 -i\sinh\varphi'(A^{(2,1)}_{TE}e^{-\frac{d\cos\theta}{2\lambdabar}}
 -
 A^{(2,2)}_{TE}e^{\frac{d\cos\theta}{2\lambdabar}})
 \nonumber\\
 =
 n\cos\theta
 (A^3_{TE}e^{i\frac{nd\cos\theta}{2\lambdabar}}
 -
 A^T_{TE}e^{-i\frac{nd\cos\theta}{2\lambdabar}}),
 \label{eqn:-d/2TE-eva}
\end{gather}
and
\begin{gather}
 \begin{pmatrix}
  A^{(2,1)}_{TM}e^{-\frac{d\sinh\varphi'}{2\lambdabar}}\\
  A^{(2,2)}_{TM}e^{\frac{d\sinh\varphi'}{2\lambdabar}}
 \end{pmatrix}
 \nonumber\\
 =\!\!
 \begin{pmatrix}
  \frac{\cos\!\theta\!+\!in\sinh\!\varphi'}{2\sinh\varphi'}&\!\!
  \frac{\cos\!\theta\!-\!in\sinh\!\varphi'}{2\sinh\varphi'}\\
  \frac{\cos\!\theta\!-\!in\sinh\!\varphi'}{2\sinh\varphi'}&\!\!
  \frac{\cos\!\theta\!+\!in\sinh\!\varphi'}{2\sinh\varphi'}
 \end{pmatrix}\!\!
 \begin{pmatrix}
   \!\!A^3_{TM}e^{i\frac{nd\!\cos\!\theta}{2\lambdabar}}\\
  \!\!A^T_{TM}e^{-i\frac{nd\!\cos\!\theta}{2\lambdabar}}
 \end{pmatrix},
 \label{eqn:A21-d/2}\\
 \begin{pmatrix}
  A^{(2,1)}_{TE}e^{-\frac{d\sinh\varphi'}{2\lambdabar}}\\
  A^{(2,2)}_{TE}e^{\frac{d\sinh\varphi'}{2\lambdabar}}
 \end{pmatrix}
 \nonumber\\
 =\!\!
 \begin{pmatrix}
   \frac{n\!\cos\!\theta\!+\!i\sinh\!\varphi'}{i2\sinh\varphi'}&\!\!
  -\frac{n\!\cos\!\theta\!-\!i\sinh\!\varphi'}{i2\sinh\varphi'}\\
  -\frac{n\!\cos\!\theta\!-\!i\sinh\!\varphi'}{i2\sinh\varphi'}&\!\!
   \frac{n\!\cos\!\theta\!+\!i\sinh\!\varphi'}{i2\sinh\varphi'}
 \end{pmatrix}\!\!
 \begin{pmatrix}
  A^3_{TE}e^{i\frac{nd\cos\theta}{2\lambdabar}}\\
  A^T_{TE}e^{-i\frac{nd\cos\theta}{2\lambdabar}}
 \end{pmatrix}.
 \label{eqn:A22-d/2}
\end{gather}
From equations (\ref{eqn:A21+d/2}), (\ref{eqn:A22+d/2}), (\ref{eqn:A21-d/2}), 
(\ref{eqn:A22-d/2}), 
we have calculated the Poynting's vectors as:
\begin{align}
 A^R_{TM}&=\frac
 {\begin{bmatrix}
   -i(\!\cos^2\!\theta\!\!+\!\!n^2\!\sinh^2\!\varphi'\!)
   \sinh(\!\frac{d\sinh\varphi'}{\lambdabar}\!)
 A^1_{TM}\\
   +2n\cos\theta\sinh\varphi'A^3_{TM}
  \end{bmatrix}}
 {\begin{bmatrix}
   2n\cos\theta\sinh\varphi'\cosh(\frac{d\sin\varphi'}{\lambdabar})\\
 +i(\cos^2\theta-n^2\sinh^2\varphi')\sinh(\frac{d\sin\varphi'}{\lambdabar})
  \end{bmatrix}}
  e^{in\frac{d\cos\theta}{\lambdabar}},\\
A^R_{TE}&=\frac
 {\begin{bmatrix}
   i(n^2\!\!\cos^2\!\theta\!\!+\!\!\sinh^2\!\!\varphi')
   \sinh(\frac{d\sinh\varphi'}{\lambdabar})
   A^1_{TE}\\
   +2n\cos\theta\sinh\varphi'A^3_{TE}
  \end{bmatrix}}
 {\begin{bmatrix}
   2n\cos\theta\sinh\varphi'\cosh(\frac{d\sin\varphi'}{\lambdabar})\\
   +i(n^2\cos^2\theta-\sinh^2\varphi')\sinh(\frac{d\sin\varphi'}{\lambdabar})
  \end{bmatrix}}
 e^{in\frac{d\cos\theta}{\lambdabar}},
\end{align}
and
\begin{align}
 A^T_{TM}&=\frac
 {\begin{bmatrix}
   2n\cos\theta\sinh\varphi'A^1_{TM}\\
   -i(\cos^2\!\theta\!\!+\!\!n^2\!\sinh^2\!\!\varphi'\!)
   \sinh(\!\frac{d\sinh\varphi'}{\lambdabar}\!)A^3_{TM}
  \end{bmatrix}}
 {\begin{bmatrix}
   2n\cos\theta\sinh\varphi'\cosh(\frac{d\sinh\varphi'}{\lambdabar})\\
   +i(\cos^2\theta-n^2\sinh^2\varphi')\sinh(\frac{d\sinh\varphi'}{\lambdabar})
  \end{bmatrix}}
 e^{in\frac{d\cos\theta}{\lambdabar}},\\
 A^T_{TE}&=\frac
 {\begin{bmatrix}
   2n\cos\theta\sinh\varphi'A^1_{TE}\\
   +i(n^2\!\!\cos^2\!\theta\!\!+\!\!\sinh^2\!\!\varphi'\!)
   \sinh(\!\frac{d\sinh\varphi'}{\lambdabar}\!\!)
   A^3_{TE}
  \end{bmatrix}}
 {\begin{bmatrix}
   2n\cos\theta\sinh\varphi'\cosh(\frac{d\sin\varphi'}{\lambdabar})\\
   +i(n^2\cos^2\theta-\sinh^2\varphi')\sinh(\frac{d\sin\varphi'}{\lambdabar})
  \end{bmatrix}}
 e^{in\frac{d\cos\theta}{\lambdabar}}.
\end{align}

We have calculated the intensity ratios by taking the long time average
as in section \ref{sec:refraction}:
\begin{align}
 \overline{R}_{TM}&=
 \frac
 {\begin{bmatrix}
   (\cos^2\!\theta\!+\!n^2\!\sinh^2\!\varphi')^2
   \sinh^2(\frac{d\omega\sinh\varphi'}{c_2})\\
   \!+\!
   4\alpha n^2\cos^2\!\theta\sinh^2\varphi'\\
   -4\sqrt{\alpha}n\sin\!\eta\cos\!\theta\sinh\!\varphi'\\
   \times\!
   (\cos^2\!\theta\!+\!n^2\!\sinh^2\!\varphi')\!
   \sinh(\frac{d\sinh\varphi'}{\lambdabar})
  \end{bmatrix}}
 {\begin{bmatrix}
   (\cos^2\!\theta\!+\!n^2\!\sinh^2\!\varphi')^2\!
   \sinh^2(\frac{d\sinh\varphi'}{\lambdabar})\\
   +\!
   4n^2\cos^2\theta\sinh^2\varphi'
  \end{bmatrix}}
 ,\label{eqn:R_TM-eva}\\
 \overline{R}_{TE}&=
 \frac
 {\begin{bmatrix}
   (\sinh^2\!\varphi'\!+\!n^2\!\cos^2\!\theta)^2\!
   \sinh^2(\frac{d\sinh\varphi'}{\lambdabar})\\
   +\!4\alpha n\sinh\!\varphi'\cos\!\theta\\
   +4\!\sin\!\eta\sqrt{\alpha}n\sinh\!\varphi'\cos\!\theta\\
   \times\!(\sinh^2\!\varphi'\!+\!n^2\!\cos^2\!\theta)^2\!
   \sinh(\frac{d\sinh\varphi'}{\lambdabar})
  \end{bmatrix}}
 {\begin{bmatrix}
   (\sinh^2\!\varphi'\!+\!n^2\!\cos\!\theta)^2\!
   \sinh^2(\frac{d\sinh\!\varphi'}{\lambdabar})\\
   +4n^2\cos^2\!\theta\sinh^2\!\varphi'
  \end{bmatrix}}.
 \label{eqn:R_TE-eva}
\end{align}

Using Snell's equation (\ref{eqn:Snell'seq})and 
the definition of $\kappa$ expressed in equation (\ref{eqn:theta-kappa}),
we have led the relations among $\kappa, \theta, n$, and $\varphi'$
expressed as:
\begin{gather}
 \frac{\sinh\varphi'}{\cos\theta}=\sqrt{(n^2-1)\kappa-1},
 \label{eqn:sinhvarphi'overcostheta-kapppa}\\
 \sinh\varphi'=\sqrt{\frac{(n^2-1)\kappa-1}{\kappa+1}},
 \label{eqn:sinhvarphi'-kapppa}
\end{gather}
Using the relations (\ref{eqn:sinhvarphi'overcostheta-kapppa}) and 
(\ref{eqn:sinhvarphi'-kapppa}), 
we have finally derived $\overline{R}_{TM}$ and $\overline{R}_{TE}$ as
functions of $d, \lambdabar, \alpha, n$, and $\kappa$ as:
\begin{align}
 \overline{R}_{TM}&=
 \frac
 {\begin{bmatrix}
   (n^2\!-1)^2(n^2\kappa\!-\!1)^2\!
   \sinh^2\bigl(\!\frac{d}{\lambdabar}
   \sqrt{\!\frac{(n^2\!-\!1)\kappa\!-\!1}{\kappa+1}}\bigr)\\
   +\!4\alpha n^2[(n^2\!-\!1)\kappa\!-\!1]\\
   -4\sqrt{\alpha}\sin\!\eta n\sqrt{(n^2\!-\!1)\kappa\!-\!1}\\
   \times\!(n^2\!-\!1)(n^2\kappa\!-\!1)\!
   \sinh\bigl(\!\frac{d}{\lambdabar}
   \sqrt{\!\frac{(n^2\!-\!1)\kappa\!-\!1}{\kappa+1}}\bigr)
  \end{bmatrix}}
 {\begin{bmatrix}
   (n^2\!-\!1)^2\!(n^2\kappa\!-\!1)^2\!
   \sinh^2\bigl(\!\!\frac{d}{\lambdabar}
   \!\sqrt{\!\frac{(n^2\!-\!1)\kappa\!-\!1}{\kappa+1}}\bigr)\\
   +4n^2[(n^2-1)\kappa-1]
  \end{bmatrix}}
 ,\label{eqn:R_TM-eva-kappa}\\
 \overline{R}_{TE}&=
 \frac
 {\begin{bmatrix}
   (n^2\!-\!1)^2(\kappa\!+\!1)^2
   \sinh^2\bigl(\!\frac{d}{\lambdabar}\!\!
   \sqrt{\!\frac{(n^2\!-\!1)\kappa\!-\!1}{\kappa+1}}\!\bigr)\\
   +4\alpha n^2[(n^2\!-\!1)\kappa\!-\!1]\\
   +4n^2[(n^2-1)\kappa-1]\\
   +4\sqrt{\alpha}\sin\!\eta n\sqrt{(n^2\!-\!1)\kappa\!-\!1}\\
   \times\!(n^2\!-\!1)(\kappa\!+\!1)\!
   \sinh\bigl(\!\frac{d}{\lambdabar}\!
   \sqrt{\frac{(n^2\!-\!1)\kappa\!-\!1}{\kappa+1}}\bigr)
  \end{bmatrix}}
 {\begin{bmatrix}
   (n^2\!-\!1)^2(\kappa\!+\!1)^2
   \sinh^2\bigl(\frac{d}{\lambdabar}\!\!
   \sqrt{\frac{(n^2-1)\kappa-1}{\kappa+1}}\bigr)\\
   +\!
   4n^2[(n^2-1)\kappa-1]
  \end{bmatrix}}
 .\label{eqn:R_TE-eva-kappa}
\end{align}

\section{Conditions for $\overline{R}=0$}\label{sec:parameter_condition}
Summing up results in sections \ref{sec:refraction} and
\ref{sec:evanescent}, 
we have attained the following equations:
\begin{align}
 \overline{R}_{TM}&=
 \frac
 {\begin{bmatrix}
   (n^2\!-1)^2(n^2\kappa\!-\!1)^2\!S_d^2(\kappa)\\
   +\!4\alpha n^2|(n^2\!-\!1)\kappa\!-\!1|\\
   -4\sqrt{\alpha}\sin\!\eta n\sqrt{|(n^2\!-\!1)\kappa\!-\!1|}\\
   \times\!(n^2\!-\!1)(n^2\kappa\!-\!1)\!S_d(\kappa)
  \end{bmatrix}}
 {\begin{bmatrix}
   (n^2\!-\!1)^2\!(n^2\kappa\!-\!1)^2\!S_d^2(\kappa)\\
   +4n^2|(n^2-1)\kappa-1|
  \end{bmatrix}}
 ,\label{eqn:R_TM-kappa}\\
 \overline{R}_{TE}&=
 \frac
 {\begin{bmatrix}
   (n^2\!-\!1)^2(\kappa\!+\!1)^2S_d^2(\kappa)\\
   +4\alpha n^2|(n^2\!-\!1)\kappa\!-\!1|\\
   +4\sqrt{\alpha}\sin\!\eta n\sqrt{|(n^2\!-\!1)\kappa\!-\!1|}\\
   \times\!(n^2\!-\!1)(\kappa\!+\!1)\!S_d(\kappa)
  \end{bmatrix}}
 {\begin{bmatrix}
   (n^2\!-\!1)^2(\kappa\!+\!1)^2S_d^2(\kappa)\\
   +\!
   4n^2|(n^2-1)\kappa-1|
  \end{bmatrix}}
 ,\label{eqn:R_TE-kappa}
\end{align}
where
\begin{equation}
 S_d(\kappa)
  :=\left\{
    \begin{matrix}
     \sin\bigl(\!\frac{d}{\lambdabar}\!
     \sqrt{\!\frac{1-(n^2\!-\!1)\kappa}{1+\kappa}}\bigr),
     &\text{if }0\le \kappa\le\frac{1}{n^2-1},\\
     \sinh\bigl(\!\frac{d}{\lambdabar}\!
     \sqrt{\!\frac{(n^2\!-\!1)\kappa\!-\!1}{\kappa+1}}\bigr),
     &\text{if }\frac{1}{n^2-1}\le \kappa
    \end{matrix}
   \right..
\end{equation}

At $\eta=\frac{\pi}{2}$ for TM mode and at $\eta=-\frac{\pi}{2}$ for
TE mode, 
we have derived equations (\ref{eqn:R_TM-kappa}) and
(\ref{eqn:R_TE-kappa}) respectively:
\begin{align}
 \overline{R}_{TM}&=
 \frac
 {\begin{bmatrix}
   \left[(n^2\!-1)(n^2\kappa\!-\!1)\!S_d(\kappa)\right.\\
   \left.
   -\!2\sqrt{\alpha}n\sqrt{|(n^2\!-\!1)\kappa\!-\!1|}\right]^2
  \end{bmatrix}}
 {\begin{bmatrix}
   (n^2\!-\!1)^2\!(n^2\kappa\!-\!1)^2\!S_d^2(\kappa)\\
   +4n^2|(n^2-1)\kappa-1|
  \end{bmatrix}}
 ,\label{eqn:R_TM-kappaAt+pi/2}\\
 \overline{R}_{TE}&=
 \frac
 {\begin{bmatrix}
   \left[(n^2\!-\!1)(\kappa\!+\!1)S_d(\kappa)
   \right.\\
   \left.
   -2\sqrt{\alpha}n\sqrt{|(n^2\!-\!1)\kappa\!-\!1|}\right]^2
  \end{bmatrix}}
 {\begin{bmatrix}
   (n^2\!-\!1)^2(\kappa\!+\!1)^2S_d^2(\kappa)\\
   +\!
   4n^2|(n^2-1)\kappa-1|
  \end{bmatrix}}
 .\label{eqn:R_TE-kappaAt-pi/2}
\end{align}

\begin{figure}[tb]
 \begin{center}
  \psfrag{kappa}{$\kappa$}
  \psfrag{alpha}{$\alpha$}
  \psfrag{d/l=pi/2}[Br][Br]{$d/\lambdabar=\frac{\pi}{2}$}
  \psfrag{d/l=pi}[Br][Br]{$d/\lambdabar=\pi$}
  \psfrag{d/l=4}[Br][Br]{$d/\lambdabar=4$}
  \psfrag{1/n^2}{$\frac{1}{n^2}$}
  \psfrag{1/(n^2-1)}{$\frac{1}{n^2-1}$}
  \includegraphics[width=70mm]{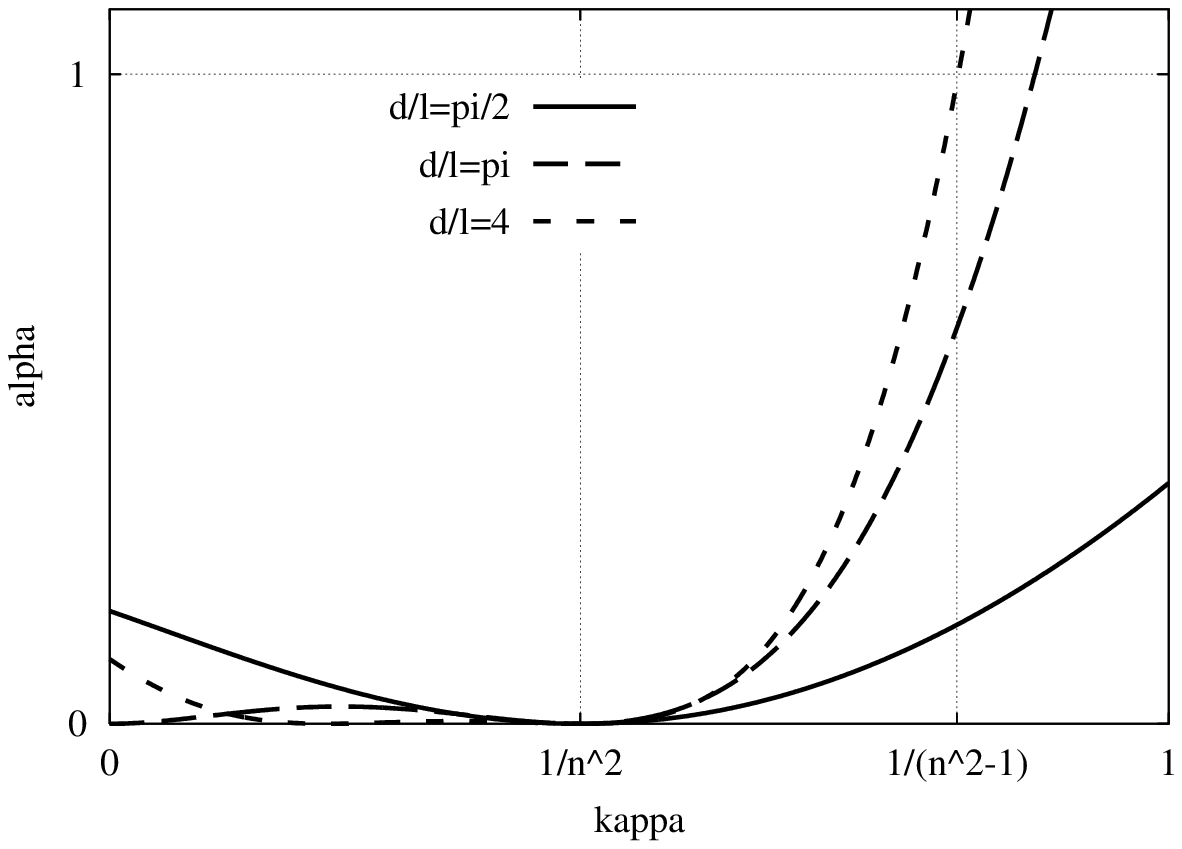}
  \caption{$\kappa-\alpha$ relation for $TM$ mode with $n=1.5$}
  \label{fig:TM-alpha-kappa-relation}
 \end{center}
\end{figure}

\begin{figure}[bt]
 \begin{center}
  \psfrag{kappa}{$\kappa$}
  \psfrag{alpha}{$\alpha$}
  \psfrag{d/l=pi/2}[Br][Br]{$d/\lambdabar=\frac{\pi}{2}$}
  \psfrag{d/l=pi}[Br][Br]{$d/\lambdabar=\pi$}
  \psfrag{d/l=4}[Br][Br]{$d/\lambdabar=4$}
  \psfrag{1/(n^2-1)}{$\frac{1}{n^2-1}$}
  \includegraphics[width=70mm]{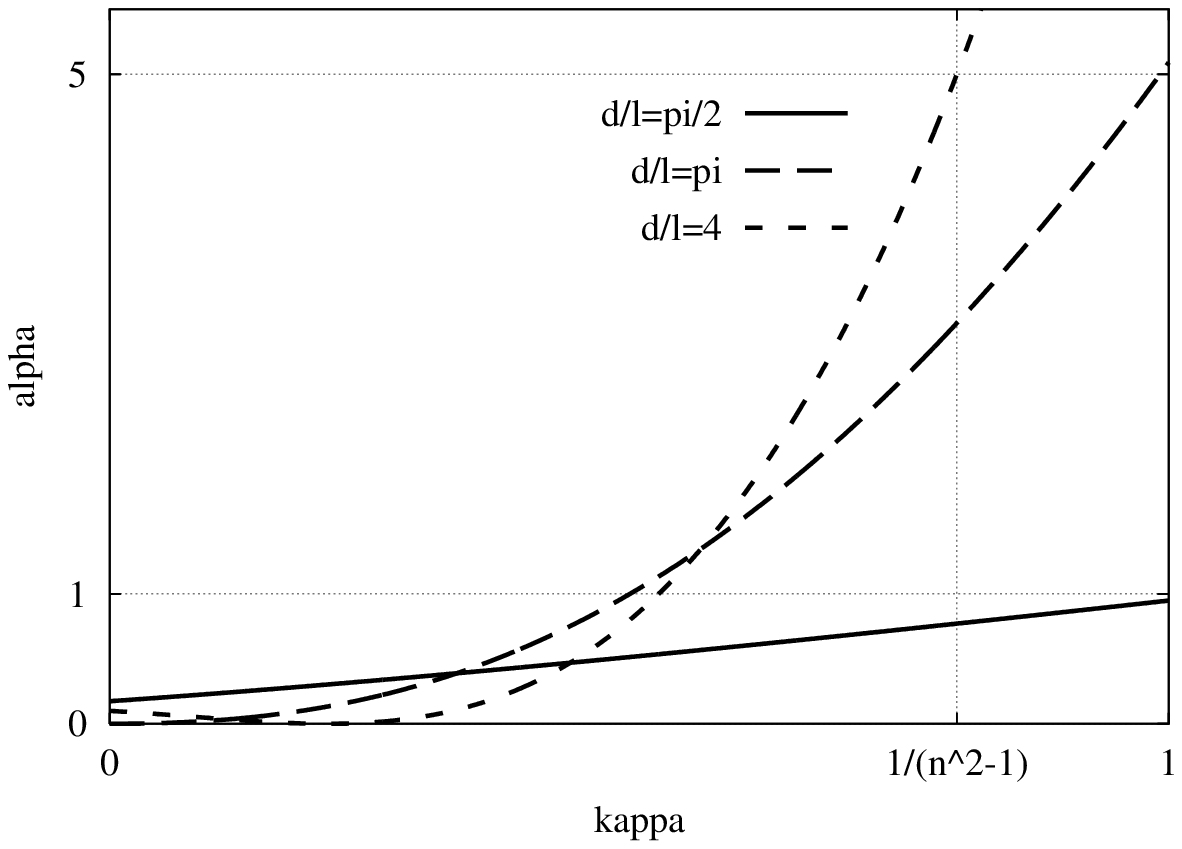}
  \caption{$\kappa-\alpha$ relation for $TE$ mode with $n=1.5$}
  \label{fig:TE-alpha-kappa-relation}
 \end{center}
\end{figure}

Because each of the numerators of (\ref{eqn:R_TM-kappaAt+pi/2}) and
(\ref{eqn:R_TE-kappaAt-pi/2}) is perfect square of a difference,
we can realize $\overline{R}_{TM}=0$ or $\overline{R}_{TE}=0$ by
selecting appropriate values for a set of free parameters 
$(\kappa, \alpha, n)$. 
For TM mode, those parameters should satisfy:
\begin{equation}
  (n^2-1)(n^2\kappa-1)S_d(\kappa)
   =2\sqrt{\alpha}n\sqrt{|(n^2-1)\kappa-1|},
   \label{eqn:kappa-alpha-TM}
\end{equation}
and for TE mode:
\begin{equation}
 (n^2-1)(\kappa+1)S_d(\kappa)=2\sqrt{\alpha}n\sqrt{|(n^2-1)\kappa-1|}.
  \label{eqn:kappa-alpha-TE}
\end{equation}
Rewriting (\ref{eqn:kappa-alpha-TM}) and (\ref{eqn:kappa-alpha-TE}):
\begin{align}
 \alpha&=\frac
 {(n^2-1)^2(n^2\kappa-1)^2S_d^2(\kappa)}
 {4n^2|(n^2-1)\kappa-1|},&TM\text{ mode},&
 \label{eqn:TM-alpha-kappa-relation}\\
 \alpha&=\frac
 {(n^2-1)^2(\kappa+1)^2S_d^2(\kappa)}
 {4n^2|(n^2-1)\kappa-1|},
 &TE\text{ mode}.&
 \label{eqn:TE-alpha-kappa-relation}
\end{align}
Equations (\ref{eqn:TM-alpha-kappa-relation}) and
(\ref{eqn:TE-alpha-kappa-relation}) are shown in
Figs.\ref{fig:TM-alpha-kappa-relation}  and
\ref{fig:TE-alpha-kappa-relation}, respectively, for $n=1.5$. 

Namely, when $\alpha$ is on either curve of
Fig.\ref{fig:TM-alpha-kappa-relation} or
Fig.\ref{fig:TE-alpha-kappa-relation}, 
$\overline{R}_{TM}$ or $\overline{R}_{TE}$ can be switched to $0$ and a
finite value depending on $\eta$ as:
\begin{align}
 \overline{R}_{TM}&=
  \left\{
   \begin{array}{cl}
    0,&\text{at }\eta=\frac{\pi}{2},\\
    \frac{4\alpha}{\alpha+1},&\text{at }\eta=-\frac{\pi}{2}
   \end{array}
   \right.,
  \label{eqn:kappa->alphaTM}\\
 \overline{R}_{TE}&=
  \left\{
   \begin{array}{cl}
    0,&\text{at }\eta=-\frac{\pi}{2},\\
    \frac{4\alpha}{\alpha+1},&\text{at }\eta=\frac{\pi}{2}
   \end{array}
   \right..
  \label{eqn:kappa->alphaTE}
\end{align}

\begin{figure}[bt]
 \begin{center}
  \psfrag{ALPHA}{$\frac{4\alpha}{\alpha+1}$}
  \psfrag{alpha}{$\alpha$}
  \includegraphics[width=70mm]{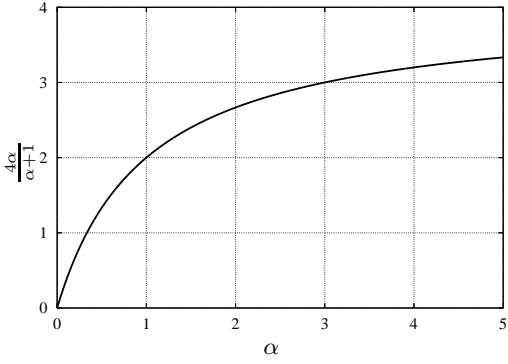}
  \caption{$\overline{R}$ at $\eta=\eta_0+\pi$}
  \label{fig:Ron}
 \end{center}
\end{figure}

The finite value of $\overline{R}$ above is shown in Fig.\ref{fig:Ron}
as a function of $\alpha$.
Equations (\ref{eqn:kappa->alphaTM}) and (\ref{eqn:kappa->alphaTE}) 
tell that when this device is coupled
with a Mach-Zehnder interferometer of Fig.\ref{fig:switch} and 
composed as an optical switch as Fig.\ref{fig:switch} using some kind of
phase controller in one of transmission lines of the interferometer, 
the intensity of the output light at 'on' does not explicitly
depend on incision angle $\theta$ nor width of vacuum layer $d$.

Finally, we will discuss on inverse functions
$\kappa=\kappa(\alpha)$ of equations (\ref{eqn:TM-alpha-kappa-relation})
and (\ref{eqn:TE-alpha-kappa-relation}). 
For equations (\ref{eqn:kappa->alphaTM}) and (\ref{eqn:kappa->alphaTE}), 
it seems more realistic to decide incision angle $\theta$ corresponding
to $\overline{R}$ after deciding the intensity ratio $\alpha$ of the
light intensity of I2 to that of I1.

For TM mode, equation (\ref{eqn:TM-alpha-kappa-relation}) shows that
$\alpha$ is a monotonic increasing function of 
$\kappa\in [\frac{1}{n^2},\infty)$ with $\alpha(\frac{1}{n^2})=0$ and
$\alpha\rightarrow\infty$ in large $\kappa$ limit,  
so that for any $\alpha\ge 0$, 
there uniquely exists $\kappa\in [\frac{1}{n^2},\infty)$ 
which satisfy equation (\ref{eqn:TM-alpha-kappa-relation}).

On the other hand, for TE mode, 
as shown by Fig.\ref{fig:TE-alpha-kappa-relation}, 
there exists some range of $\alpha$ value which cannot be reached for
any $\kappa$ value. 
Actually, equation (\ref{eqn:TE-alpha-kappa-relation}) becomes around
$\kappa=0$,
\begin{align}
 \alpha
 &=\frac{(n^2-1)^2}{4n^2}\sin^2\bigl(\frac{d}{\lambdabar}\bigr)
 \nonumber\\
 &+\!\!\frac{(n^2\!\!-\!\!1)^2\!(n^2\!\!+\!\!1)^2}{8n^2}
 \!\sin\bigl(\!\frac{2d}{\lambdabar}\!\bigr)
 \!\bigl\{\!\tan\bigl(\!\frac{d}{\lambdabar}\!\bigr)\!
 -\!\frac{n^2}{n^2\!+\!1}\frac{d}{\lambdabar}\!\bigr\}\kappa
 \nonumber\\
 &
 +O(\kappa^2).
 \label{eqn:TE-taylor}
\end{align}

The coefficient of the first order of $\kappa$ in equation
(\ref{eqn:TE-taylor}) is positive for a region 
$0\le\frac{d}{\lambdabar}<\pi$, and within this region, 
$\alpha$ is a monotonic increasing function of $\kappa$ around
$\kappa=0$. 
Therefore, in the region $0\le\frac{d}{\lambdabar}<\pi$, 
the $\alpha$ takes the minimum value
$\frac{(n^2-1)^2}{4n^2}\sin^2\bigl(\frac{d}{\lambdabar}\bigr)$
at $\kappa=0$. 
For the region $\pi\le\frac{d}{\lambdabar}$, 
since $\alpha=0$ at 
$\kappa=\frac{d-\pi\lambdabar}{dn^2+\pi\lambdabar-d}$,
there exists at least one 
$\kappa\ge\frac{d-\pi\lambdabar}{dn^2+\pi\lambdabar-d}$ for any
$\alpha\ge 0$.

Based on those discussion results, 
we have attained the following statements for $\alpha$ value for
$\overline{R}_{TE}=0$ in TE mode:
\begin{itemize}
 \item for $0\le\frac{d}{\lambdabar}<\pi$:\\
       There exists a $\kappa$ for $\overline{R}_{TE}=0$ for any
       non-negative value of $\alpha$ larger than 
       $\frac{(n^2-1)^2}{4n^2}\sin^2\bigl(\frac{d}{\lambdabar}\bigr)$,
       and
 \item for $\pi\le\frac{d}{\lambdabar}$:\\
       there exists $\kappa$ for $\overline{R}_{TE}=0$ for any non-negative
       value of $\alpha$.
\end{itemize}

\section{Conclusion}\label{sec:result}
We have taken as the study model an infinite dielectric region with
refractive index $n$ divided by an two dimensionally infinite vacuum
layer between $z=\frac{d}{2}$ and $z=-\frac{d}{2}$. 
Two input lights are injected from the upper region 1
($z>\frac{d}{2}$) and the lower region ($z<-\frac{d}{2}$) with intensity
of 1 and $\alpha$, respectively, and with the same injection angle
$\theta$. 
We have solved the intensity of output lights into region 1 and 3
directly from the Maxwell's equations. 
As the results, we have found that one of two output lights can be made
zero by appropriately selecting set of values of $\alpha$, phase
difference of the two input lights $\eta$, $d$ and $\theta$. 
Namely,
\begin{enumerate}
 \item There exists a set of parameters 
       $(\alpha_0, \theta_0, d_0, \eta_0)$
       that makes one of output light zero ($\overline{R}=0$).
       And for these set of values and replacing $\eta_0$ to
       $\eta_0+\pi$, 
       $$\overline{R}=\frac{4\alpha}{\alpha+1}.$$
       Here $\eta_0$ is $\frac{\pi}{2}$ for TM mode and
       $-\frac{\pi}{2}$ for TE mode,
 \item When a TM mode light is used as the input light, 
       for any set of $(\theta, d)$, 
       a value of $\alpha$ can be obtained by equation 
       (\ref{eqn:TM-alpha-kappa-relation}) which makes $\overline{R}=0$
       at $\eta=\frac{\pi}{2}$. 
       Particularly for $\theta$ larger than the critical angle and
       evanescent light exist in region 2 (vacuum region between upper
       and lower dielectric region), $\alpha$ for $\overline{R}=0$
       satisfies
       $\alpha\ge\frac{(n^2-1)^2}{4n^4}(\frac{d}{\lambdabar})^2$,
       and
 \item When a TE mode light is used as the input light, for any set of
       $(\theta, d)$, 
       a value of $\alpha$ can be obtained by equation
       (\ref{eqn:TE-alpha-kappa-relation}) 
       which makes $\overline{R}=0$ at $\eta=-\frac{\pi}{2}$. 
       Particularly for $\theta$ larger than the critical angle and
       evanescent light exist in region 2, 
       $\alpha$ for $\overline{R}=0$ satisfies
       $\alpha\ge\frac{n^2-1}{4}(\frac{d}{\lambdabar})^2$.
\end{enumerate}

The above results 1-3 suggests a possibility that, using a Mach-Zehnder
interferometer with controlling phase difference between two transmission
lines to $\frac{\pi}{2}$ or $-\frac{\pi}{2}$, 
we can control the output light on and off. 
One point is that the output light is perfectly eliminated at switched 'off'
without depending on $\alpha$. 
This tells that the new switch proposed here is better than the usual
Mach-Zehnder interferometer because in a Mach-Zehnder interferometer, a
perfect branching of input light of 50 : 50 is necessary to get the
output zero at 'off'. 
More notable point better than a Mach-Zehnder interferometer is
that, the output light is controlled to on and off with a light much
weaker than the output light at 'on'. 
In a case of $\alpha=0.10$ for example, the output light intensity at
'on' is $\frac{4\alpha}{\alpha+1}=0.36$. 
In this case, the output light is controlled by a control light with
intensity of $\frac{1}{3.6}$ of the output light. 
From engineering point of view, this means that a control light can be
divided to several optical switches which is one of critically important
features for the switch to be applied in logic circuits.

Since a real optical device has a finite size of several to ten times
larger than the light wavelength whereas an infinite model is studied in
this paper, 
effects of finite size such as the effect of the higher modes must be
investigated in future. 
Furthermore, effect of real boundary plane being not perfectly flat and
not perfectly parallel shall also be investigated.

\begin{acknowledgments}
The research is supported by the science research promotion fund 2006 \&
2007 from The Promotion and Mutual Aid Corporation for Private Schools
of Japan.
\end{acknowledgments}


\end{document}